\documentclass[showpacs,preprint,superscriptaddress,amssymb]{revtex4}
\usepackage{graphicx}
\begin{document}

\title{Thermodynamic duality between RN black hole and 2D dilaton gravity}

\author{Yun Soo Myung\footnote{ysmyung@inje.ac.kr}}
\affiliation{Institute of Mathematical Science and
  School of Computer Aided Science, Inje University, Gimhae 621-749,
  Korea}
\author{Yong-Wan Kim\footnote{ywkim65@gmail.com}}
\affiliation{Institute of Mathematical Science and
  School of Computer Aided Science, Inje University, Gimhae 621-749,
  Korea}
\author{Young-Jai Park\footnote{yjpark@sogang.ac.kr}}
\affiliation{Department of Physics  and Center for Quantum
  Spacetime, Sogang University, Seoul 121-742, Korea}

\begin{abstract}
All thermodynamic quantities of the Reissner-Nordstr\"om (RN) black
hole can be obtained from  the dilaton and its potential of two
dimensional (2D) dilaton gravity. The dual relations of four
thermodynamic laws are also established. Furthermore, the
near-horizon thermodynamics of the extremal RN black hole is
completely described by the Jackiw-Teitelboim theory which is
obtained by perturbing around the AdS$_2$-horizon.

\end{abstract}


\maketitle


\section{ Introduction}

Since the pioneering work of Bekenstein \cite{bek,bek2,bek3} and Hawking
\cite{haw}, which proved that the entropy of black hole is
proportional to the area of its horizon in the early 1970s, the
research of the black hole  thermodynamics has greatly improved.
Especially, the proof of the Hawking radiation stimulated the
enthusiasm for studying the thermal property of the black hole. It
was found that if the surface gravity of the black hole is
considered to be the temperature and the outer horizon area is
considered to be the entropy, four laws of the black hole
thermodynamics, which correspond with the four laws of the
elementary thermodynamics, has been established~\cite{BCH}.

On the other hand, 2D dilaton gravity
has been used in various situations as an effective
description of 4D gravity after a black hole in string theory has
appeared \cite{witten,wit1}. Hawking radiation and thermodynamics of
this black hole has been analyzed by several authors \cite{crff,crf1,crf2,crf3}.
Another 2D theories, which were originated from the
Jackiw-Teitelboim (JT) theory \cite{jackiw,Teit}, have been also
studied \cite{JT-theories,JT1,JT2}. Although in this theory the curvature
is constant and negative, it has a black hole solution, which
implies a non-trivial causal structure and in turn generates
interesting non-trivial thermodynamics \cite{cacl,cac2,cac3,cac4,cac5,cac6}. Moreover,
Fabbri {\it et. al.} \cite{fnn} partially demonstrated the duality
of the thermodynamics between near-extremal RN black hole and the
JT theory because they considered temperature and entropy only.
Recently, Grumiller and McNees ~\cite{GM} discussed black hole thermodynamics
in the 2D dilaton gravity (for a review see \cite{gkv}).

 In this Letter, we completely describe
the thermodynamic duality between the RN black hole and the 2D
dilaton gravity. The key ingredient is to fully use the dilaton
potential induced by the dimensional reduction and the conformal
transformation. Moreover, we also show the thermodynamic duality
between near-extremal RN black hole and the JT theory.

\section{ RN thermodynamics}

Consider the RN black hole, whose metric is given by
\begin{equation} \label{MF}
ds^2_{RN}=-U(r)dt^2+U^{-1}(r)dr^2+r^2d\theta^2+r^2\sin^2\theta
d\varphi^2
\end{equation}
with $U(r)=1-2M/r+Q^2/r^2$. Here, $M$ and $Q$ are the mass and the
electric charge of the RN black hole, respectively. Then, the
inner ($r_-$) and the outer ($r_+$) horizons are obtained as
$r_\pm=M\pm\sqrt{M^2-Q^2}$, which satisfy $U(r_\pm)=0$. For
$M_e=Q$, we have an extremal RN black hole at $r_e=Q$. In this work we
consider the case of fixed charge $Q$ for simplicity~\cite{CEJM}.
The other case of the fixed potential $\Phi=Q/r_+$ will have
parallel with the fixed charge case.

For the RN black hole, the relevant thermodynamic quantities are
given by the Bekenstein-Hawking entropy and Hawking temperature
\begin{eqnarray} \label{as}
S_{BH}(M,Q)&=& \pi \Big(M+\sqrt{M^2-Q^2}\Big)^2,\\
T_{H}(M,Q)&=& \frac{\sqrt{M^2-Q^2}}{2\pi(M+\sqrt{M^2-Q^2})^2}.
\label{at}
\end{eqnarray}
Then, using the Eqs. (\ref{as}) and (\ref{at}), the heat capacity
$C=(dM/dT_{H})_Q$ and Helmholtz free energy $F$ are obtained to be
\begin{eqnarray}\label{ac}
C(M,Q)&=&\frac{2\pi
\sqrt{M^2-Q^2}(M+\sqrt{M^2-Q^2})^3}{2Q^2-M(M+\sqrt{M^2-Q^2})}, \\
\label{af} F(M,Q)&=&E-T_{H}S_{BH}=E-\frac{1}{2} \sqrt{M^2-Q^2}
\end{eqnarray}
with $E=M-Q$. We note here that one has to use the extremal black
hole as  background  for fixed charge ensemble~\cite{CEJM}.
\begin{figure*}[t!]
   \centering
   \includegraphics{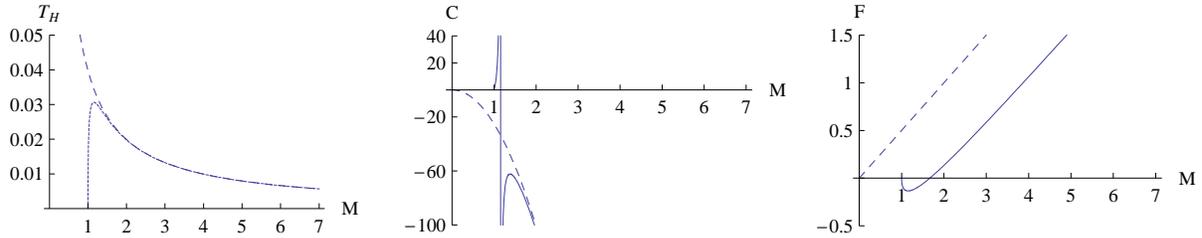}
\caption{Thermodynamic quantities of temperature, heat capacity, and
free energy as function of mass $M$ with fixed $Q=1$. The solid
curves represent the RN black hole, while the dashed curves denote
the Schwarzschild black hole with $Q=0$.} \label{fig1}
\end{figure*}

As is shown in Fig. 1, the features of thermodynamic quantities
are as follows: First of all,  the whole region splits into the
near-horizon phase of $Q<M<M_m$ with $M_m=\frac{2}{\sqrt{3}}Q$ and
Schwarzschild phase of $M>M_m$.  i) The temperature is zero
($T_H=0$)  at $M=Q$, maximum $T_m=\frac{1}{6\sqrt{3} \pi Q}$ at
$M=M_m$, and for $M>M_m$, it shows the Schwarzschild behavior. ii)
The heat capacity $C$ determines thermodynamic stability. For
$Q<M<M_m$, it is stable because of $C>0$, while for $M>M_m$, it is
unstable ($C<0$) as is shown in the Schwarzschild case. Here, we
have $C=0$ at  $M=Q$, and importantly $C$ blows up at $M=M_m$.
iii) The free energy is an important quantity to determine the
presumed phase transition. The free energy  is always negative and
it is decreasing ($E<T_HS_{BH}$) for the near-horizon phase and
increasing ($E>T_HS_{BH}$) for the Schwarzschild phase. It takes
the minimum value $F_m=\frac{Q(\sqrt{3}-2)}{2}$ at $M=M_m$ and
$F=0$ at $M=Q$. Here, we classify the extremal RN black hole by
the conditions of $T_H=0,~C=0,~ S_{BH}=\pi Q^2,~F=0$.

\section{2D dilaton gravity induced by dimensional reduction}

We start with the four-dimensional action whose
solution is given by Eq.(\ref{MF}) as
\begin{equation}
\label{action} I=\frac{1}{16\pi}\int d^4x
          \sqrt{-g}[R-F_{\mu\nu}F^{\mu\nu}]
\end{equation}
with $F_{\mu\nu}F^{\mu\nu}=2Q^2/r^2$. After the dimensional
reduction by integrating over $S^2$ and eliminating the kinetic
term by using the conformal transformation
\begin{equation} \label{conft}
\bar{g}_{\mu\nu}=\sqrt{\phi}~g_{\mu\nu},~~~\phi=\frac{r^2}{4},
\end{equation}
the reparameterized action is obtained as
\begin{eqnarray}\label{repara-action}
\bar{I}^{(2)}=\int d^2x \sqrt{-\bar{g}}[\phi\bar{R}_2+V(\phi)],
\end{eqnarray}
where the 2D Ricci scalar $\bar{R}_2=-\frac{U''}{\sqrt{\phi}}$ and
the dilaton potential is given by
\begin{equation}
V(\phi)=\frac{1}{2\sqrt{\phi}}-\frac{Q^2}{8\phi^{3/2}}.
\label{pot}
\end{equation}
This is an effective 2D dilaton gravity with $G_2=1/2$
~\cite{jackiw}. The two equations of motion are
\begin{eqnarray} \label{newat1}
\nabla^2\phi=V(\phi),\\
\bar{R}_2=-V'(\phi),  \label{newat2}
\end{eqnarray}
where the derivative of $V$ is given by
\begin{equation}
V'(\phi)=-\frac{1}{4\phi^{3/2}}+\frac{3Q^2}{16\phi^{5/2}}.
\end{equation} Then, we obtain the general solution to Eqs.
(\ref{newat1}) and (\ref{newat2}) by choosing a conformal gauge of
$\bar{g}_{tx}=0$ \cite{GKL,cfnn,mkp} as
\begin{eqnarray}
\frac{d\phi}{dx}&=&2(J(\phi)-{\cal C}), \\
 ds^2_{2D}&=&-(J(\phi)-{\cal C})dt^2+\frac{dx^2}{J(\phi)-{\cal C}},
\end{eqnarray}
where $J(\phi)$ is the integration of $V$ given by
\begin{equation}
J(\phi)=\int^{\phi}V(\tilde{\phi})d\tilde{\phi}=\sqrt{\phi}+\frac{Q^2}{4\sqrt{\phi}}~.
\end{equation}
Here, ${\cal C}$ is a coordinate-invariant constant of
integration, which is identified with the mass $M$ of the RN black
hole. We note here the important connection between $J(\phi)$ and
the metric function $U(r(\phi))$ with $r=2\sqrt{\phi}$:
$\sqrt{\phi}~U(\phi)=J(\phi)-M$. For $\phi={\rm const.}$, we have
the AdS$_2$-horizon which satisfies $J=M$ that is equivalent to
$U(r_\pm)=0$.

\begin{figure}[t!]
   \centering
   \includegraphics{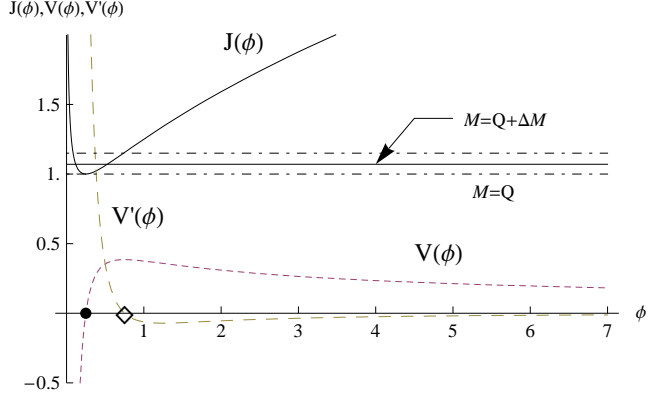}
\caption{Three graphs for the 2D dilaton gravity with $Q=1$. The
solid curve describes $J(\phi)$, the dotted curve gives $V(\phi)$,
the dashed curve denotes $V'(\phi)$, and the near-horizon region
is between the dot-dashed lines. $\bullet$($\diamond$) represent
$\phi=\phi_0$($\phi_m$). The horizontal line at $M=Q+\Delta M$ is
for the mass perturbation of the Eq. (23).} \label{fig2}
\end{figure}

As is shown in Fig. 2, we observe two important points: one
($\bullet$) is $\phi=\phi_0=Q^2/4$, where $J=Q,~J'=V=0,
~J''=V'=4/Q^3$. Another ($\diamond$) is $\phi=\phi_m=3Q^2/4$, where
$J=\frac{2}{\sqrt{3}}Q,~J'=V=\frac{2}{3\sqrt{3}}Q,~J''=V'=0$. The
former shows the extremal configuration, while the latter indicates
the unstable maximum point.
 All thermodynamic quantities can be explicitly expressed in terms of  the dilaton $\phi$, the dilaton
 potential $V(\phi)$, its integration $J(\phi)$, and its derivative
 $V'(\phi)$
as
 \begin{eqnarray} \label{phitds}
 &{}&S_{BH}(\phi)= 4\pi \phi,~T_{H}(\phi)=\frac{V(\phi)}{4\pi},~C(\phi) = 4\pi \frac{V(\phi)}{V'(\phi)} \nonumber \\
 &{}&F(\phi)=J(\phi)-J(\phi_0)-\phi V(\phi).
\end{eqnarray}
These are our main results.
\begin{figure*}[t!]
   \centering
   \includegraphics{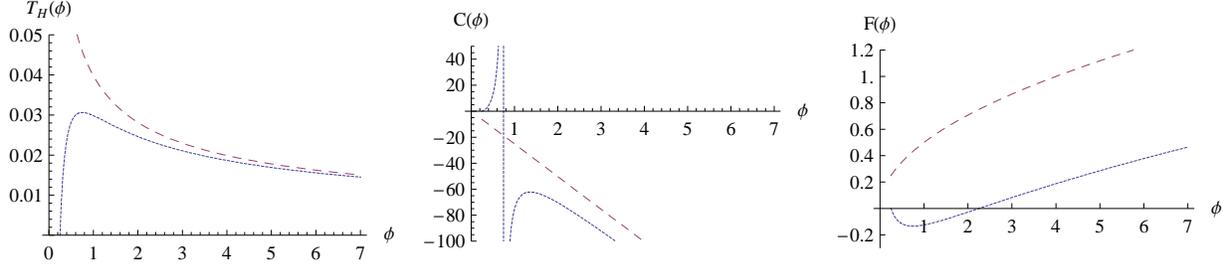}
\caption{Graphs for RN thermodynamic quantities expressed in terms
of $J,V,V'$ as functions of $\phi$. Here $\phi$ plays the role of
the entropy. The solid curves represent the RN black hole with
$Q=1$, while the dashed curves denote the Schwarzschild black hole
with $Q=0$. } \label{fig3}
\end{figure*}
In Fig. 3, we have the corresponding dual graphs, which are nearly
the same as in Fig. 1. For $\phi_0<\phi<\phi_m$, we have the JT
phase, whereas for $\phi>\phi_m$, we have the Schwarzschild phase.
At the extremal point $M=Q$, we have $T_{H}=0,~C=0,~F=0$, which
are determined by $V(\phi_0)=0$. On the other hand, at the maximum
point $M=M_m$, one has $T_{H}=T_m,~C=\pm \infty,~F=F_{m}$, which
are fixed by $V'(\phi_m)=0$.

\section{ Four thermodynamic laws}

As is shown in Table I, we confirm that the dual relation of four
thermodynamic laws can be easily realized by the 2D dilaton and
its potential. The first law is given by the definition of $J$:
$dJ=Vd\phi$.  Concerning on the third law, it has a rather
different status from the others. It was proved that an infinite
amount of time is required for the near-extremal RN black hole to
decay to extremality at $T_H=0$~\cite{fnn}. Similarly, we require
that it take an infinite time to arrive at $V(\phi_0)=0$ from
$V(\phi_0)\not=0$ located at the near-horizon phase.

\begin{table*}
\begin{tabular}{|c|c|c|c|c|}
  \hline
   & 0th law & 1st law & 2nd law & 3rd law \\
  \hline
   RN black hole& $T_H=\kappa/2 \pi$ & $dM=T_H dS_{BH}+(\Phi-\Phi_e)dQ$ & $\Delta S_{BH} \ge 0$ & $T_{H} \to 0$
   \\\hline
   2D dilaton gravity& $T_H=V/4\pi$  & $dJ=Vd\phi+(\frac{Q}{2\sqrt{\phi}}-\frac{Q}{2\sqrt{\phi_0}})dQ$ & $\Delta \phi \ge 0$ & $V \to 0$ \\
  \hline
\end{tabular}

\label{table I}
  \caption{Dual relations of four thermodynamic laws. Here the potential and its extremal value are given by
  $\Phi=Q/r_+$ and $\Phi_e=Q/r_e=1$~\cite{CEJM}. }
\end{table*}

\section{ Duality between near-extremal RN black hole and JT theory}

It is very interesting to explore the near-horizon thermodynamics
to the extremal RN black hole. This could be obtained by inserting
$\Delta M=M-Q$ into Eqs. (\ref{as})-(\ref{af}) to leading order in
$\sqrt{\Delta M}$. Then, all near-horizon thermodynamic quantities
are given by
\begin{eqnarray}
S_{NH}&=&2 \pi Q \sqrt{2Q\Delta M},~
 T_{NH}= \frac{\sqrt{2 Q\Delta M}}{2\pi
Q^2},\\
  C_{NH}&=& 2 \pi Q\sqrt{2Q\Delta
M}, ~F_{NH}= -\frac{\sqrt{2Q\Delta M}}{2}
\end{eqnarray}
with the definition of $f_{NH} \equiv f(M,Q)-f(Q,Q)$. In general,
the near-horizon quantities do not satisfy the first law of
thermodynamics. Instead, we find $2\Delta M=T_{NH}S_{NH}$ and
$S_{NH} = C_{NH}$, which show the same bebavior as in the
non-rotating BTZ black hole~\cite{myung}.

In order to find the  AdS$_2$ gravity of the JT theory, we
consider the perturbation around the  AdS$_2$-horizon as
\begin{eqnarray}
J(\phi)&\simeq& Q+\frac{V'(\phi_0)}{2} \varphi^2,\\
M&\simeq&Q[1+ k \alpha^2]\equiv Q+\Delta M \end{eqnarray} with
$\varphi=\phi-\phi_0$, $J'(\phi_0)=V(\phi_0)=0$, and
$J''(\phi_0)=V'(\phi_0)$. Introducing the new coordinates
\begin{equation} \label{coord}
\tilde{t}=\alpha t,~\tilde{x}=\frac{x-x_e}{\alpha},
\end{equation}
then the perturbed metric function with the perturbed dilaton
$\varphi = \alpha\tilde{x}$ is given by
\begin{eqnarray}
 ds^2_{JT}=-\Big[\frac{V'(\phi_0)}{2}
    \tilde{x}^2-kQ\Big]d\tilde{t}^2
  + \frac{d\tilde{x}^2}{\Big[\frac{V'(\phi_0)}{2}\tilde{x}^2-kQ\Big]}, \nonumber\\
 \label{metricads}
\end{eqnarray}
which shows a locally AdS$_2$ spacetime. If $k=0$, it is a globally
AdS$_2$ spacetime. Moreover, the mass deviation $\Delta M$ is the
conserved parameter of the JT theory \cite{cfnn}.

Now, we are in a position to
describe the near-horizon thermodynamics from the JT theory. From
the null condition of the metric function in Eq.
(\ref{metricads}), we have the positive root
\begin{equation}
\tilde{x}_+=\sqrt{\frac{2kQ}{V'(\phi_0)}} \to
\varphi_+=\frac{Q}{2}\sqrt{2Q \Delta M}.
\end{equation}
Then, the JT thermodynamics can be obtained by perturbing the
thermodynamic quantities in Eq. (\ref{phitds}) around $\phi=\phi_0$
as
\begin{eqnarray}
S_{JT}&=&4\pi\varphi_+,~
 T_{JT}=\frac{V'(\phi_0)\varphi_+}{4\pi},\\
C_{JT}&=& 4\pi \varphi_+,~
 F_{JT}= -\frac{\varphi_+}{Q}.
\end{eqnarray}
As a result, we confirm that $f_{JT}=f_{NH}$. This means that the
near-horizon (AdS$_2$) thermodynamics of the extremal RN black hole
can be completely described by the JT theory. We note that the
equality between the near-horizon thermodynamics and JT theory was
partially proved for the entropy and temperature up to
now~\cite{fnn}. In this work, we have completely described the dual
relations for all thermodynamic quantities including the heat
capacity and free energy. Furthermore, assuming  asymptotic symmetry
at boundary with a periodicity of $2\pi\beta$ in $\tilde{t}$
\cite{NSN}, we could also show that the JT entropy $S_{JT}$ is equal
to the statistical entropy
\begin{equation}
S_{CFT_1}=2\pi \sqrt{\frac{cL_0}{6}}=2\pi Q \sqrt{2Q\Delta M}
\end{equation}
with the central charge $c=12Q^3\alpha/\beta$ and the Virasoro
generator $L_0^R=kQ\alpha \beta$. This is a realization of
 the AdS$_2$/CFT$_1$ correspondence in the near-horizon
region.

\section{ Discussions}

For  the Schwarzschild black hole with $Q=0$, we have no point of
$\phi=\phi_0,~\phi_m$. Here we have $T_H=1/8\pi M,~C=-8\pi M^2,~F=
M/2$ in the Schwarzschild phase. Similarly, we have $J=
\sqrt{\phi},~V=1/2\sqrt{\phi},~V'=-1/4\phi^{3/2}$ in the 2D
dilaton gravity approach. Considering $\sqrt{\phi}= M$ in the
Schwarzschild phase, we confirm that the thermodynamic duality
between the Schwarzschild black hole and 2D dilaton gravity holds.
Also it is clear that the duality is manifested by the conformal
transformation (\ref{conft}) after integration over $S^2$ because
it enforces the dilaton potential to be Eq. (\ref{pot}).

In conclusion, we have completely described the thermodynamic
duality between the RN black hole and the 2D dilaton gravity. The
key ingredient is to fully use the dilaton potential induced by
the dimensional reduction and the conformal transformation. Even
though, for simplicity, we have considered the RN black hole and
the Schwarzschild black hole with $Q=0$, the dilaton potential is
available for all spherically symmetric black holes in four
dimensions. Hence, our approach simplifies the study on the black
hole significantly such that the 4D black hole thermodynamics can
be completely described by the 2D dilaton potential.

\section*{Acknowledgement}
Two of us (Y.S. Myung and Y.-J. Park) were supported by the
Science Research Center Program of the Korea Science and
Engineering Foundation through the Center for Quantum Spacetime of
Sogang University with grant number R11-2005-021.  Y.-W. Kim was
supported by the Korea Research Foundation Grant funded by Korea
Government (MOEHRD): KRF-2007-359-C00007.

\end{document}